\documentclass{emulateapj}
\usepackage{apjfonts}
\usepackage{rotating}

\newcommand{\pivec}{\mbox{\boldmath $\pi$}}

\lefthead{HWANG ET AL.} 
\righthead{MICROLENSING SHORT-TERM ANOMALY}

\begin{document}

\title{INTERPRETATION OF A SHORT-TERM ANOMALY IN THE GRAVITATIONAL MICROLENSING EVENT MOA-2012-BLG-486}

\author{
K.-H. Hwang$^{1}$,
J.-Y. Choi$^{1}$,
I. A. Bond$^{2,43}$,
T. Sumi$^{3,43}$,
C. Han$^{1,44,47}$,
B. S. Gaudi$^{4,44}$,
A. Gould$^{4,44}$,
V. Bozza$^{5,6}$,
J.-P. Beaulieu$^{7,45}$,
Y. Tsapras$^{8,9,46}$
\\and\\
F. Abe$^{10}$,
D. P. Bennett$^{11}$,
C. S. Botzler$^{12}$,
P. Chote$^{13}$,
M. Freeman$^{12}$,
A. Fukui$^{14}$,
D. Fukunaga$^{10}$,
P. Harris$^{13}$,
Y. Itow$^{10}$,
N. Koshimoto$^{3}$,
C. H. Ling$^{2}$,
K. Masuda$^{10}$,
Y. Matsubara$^{10}$,
Y. Muraki$^{10}$,
S. Namba$^{3}$,
K. Ohnishi$^{15}$,
N. J. Rattenbury$^{12}$,
To. Saito$^{16}$,
D. J. Sullivan$^{13}$,
W. L. Sweatman$^{2}$,
D. Suzuki$^{3}$,
P. J. Tristram$^{17}$,
K. Wada$^{3}$,
N. Yamai$^{18}$,
P. C. M. Yock$^{12}$,
A. Yonehara$^{18}$
\\(The MOA Collaboration)\\
L. Andrade de Almeida$^{19,20}$,
D. L. DePoy$^{21}$,
Subo Dong$^{22}$,
F. Jablonski$^{19}$,
Y. K. Jung$^{1}$,
A. Kavka$^{4}$,
C.-U. Lee$^{23}$,
H. Park$^{1}$,
R. W. Pogge$^{4}$,
I.-G. Shin$^{1}$,
J. C. Yee$^{4}$
\\(The $\mu$FUN Collaboration)\\
M. D. Albrow$^{24}$,
E. Bachelet$^{25,26}$,
V. Batista$^{7}$,
S. Brillant$^{27}$,
J. A. R. Caldwell$^{28}$,
A. Cassan$^{7}$,
A. Cole$^{29}$,
E. Corrales$^{7}$,
Ch. Coutures$^{7}$,
S. Dieters$^{25}$,
D. Dominis Prester$^{30}$,
J. Donatowicz$^{31}$,
P. Fouqu\'{e}$^{25,26}$,
J. Greenhill$^{29}$,
U. G. J{\o}rgensen$^{32}$,
S. R. Kane$^{33}$,
D. Kubas$^{7,27}$,
J.-B. Marquette$^{7}$,
R. Martin$^{34}$,
P. Meintjes$^{35}$,
J. Menzies$^{36}$,
K. R. Pollard$^{24}$,
A. Williams$^{37}$,
D. Wouters$^{7}$
\\(The PLANET Collaboration)\\
D. M. Bramich$^{38}$,
M. Dominik$^{39,45}$,
K. Horne$^{39,45}$,
P. Browne$^{39}$,
M. Hundertmark$^{39}$,
S. Ipatov$^{40}$,
N. Kains$^{38}$,
C. Snodgrass$^{41}$,
I. A. Steele$^{42}$,
R. A. Street$^{8}$
\\(The RoboNet Collaboration)
}

\affil{$^{1}$Department of Physics, Chungbuk National University, Cheongju 361-763, Republic of Korea}
\affil{$^{2}$Institute of Information and Mathematical Sciences, Massey University, Private Bag 102-904, North Shore Mail Centre, Auckland, New Zealand}
\affil{$^{3}$Department of Earth and Space Science, Osaka University, Osaka 560-0043, Japan}
\affil{$^{4}$Department of Astronomy, The Ohio State University, 140 West 18th Avenue, Columbus, OH 43210, USA}
\affil{$^{5}$Dipartimento di Fisica ``E. R. Caianiello,'' Universit\`a degli Studi di Salerno, Via S. Allende, I-84081 Baronissi (SA), Italy}
\affil{$^{6}$Instituto Nazionale di Fisica Nucleare, Sezione di Napoli, Italy}
\affil{$^{7}$UPMC-CNRS, UMR7095, Institut d'Astrophysique de Paris, 98bis boulevard Arago, 75014 Paris, France}
\affil{$^{8}$Las Cumbres Observatory Global Telescope Network, 6740B Cortona Dr, Goleta, CA 93117, USA}
\affil{$^{9}$School of Physics and Astronomy, Queen Mary University of London, Mile End Road, London E1 4NS, UK}
\affil{$^{10}$Solar-Terrestrial Environment Laboratory, Nagoya University, Nagoya, 464-8601, Japan}
\affil{$^{11}$University of Notre Dame, Department of Physics, 225 Nieuwland Science Hall, Notre Dame, IN 46556-5670, USA}
\affil{$^{12}$Department of Physics, University of Auckland, Private Bag 92-019, Auckland 1001, New Zealand}
\affil{$^{13}$School of Chemical and Physical Sciences, Victoria University, Wellington, New Zealand}
\affil{$^{14}$Okayama Astrophysical Observatory, National Astronomical Observatory of Japan, Asakuchi, Okayama 719-0232, Japan}
\affil{$^{15}$Nagano National College of Technology, Nagano 381-8550, Japan}
\affil{$^{16}$Tokyo Metropolitan College of Aeronautics, Tokyo 116-8523, Japan}
\affil{$^{17}$Mt. John University Observatory, P.O. Box 56, Lake Tekapo 8770, New Zealand}
\affil{$^{18}$Department of Physics, Faculty of Science, Kyoto Sangyo University, 603-8555, Kyoto, Japan}
\affil{$^{19}$Instituto Nacional de Pesquisas Espaciais, S\~ao Jos\'e dos Campos, SP, Brazil}
\affil{$^{20}$Instituto de Astronomia, Geof\'isica e Ci\^encias Atmosf\'ericas - IAG/USP Rua do Mat\~ao, 1226, Cidade Universit\'aria, S\~ao Paulo-SP - Brasil}
\affil{$^{21}$Department of Physics and Astronomy, Texas A\&M University, College Station, TX 77843, USA}
\affil{$^{22}$Institute for Advanced Study, Einstein Drive, Princeton, NJ 08540, USA}
\affil{$^{23}$Korea Astronomy and Space Science Institute, Daejeon 305-348, Republic of Korea}
\affil{$^{24}$University of Canterbury, Dept. of Physics and Astronomy, Private Bag 4800, 8020 Christchurch, New Zealand}
\affil{$^{25}$Universit\'{e} de Toulouse, UPS-OMP, IRAP, 31400 Toulouse, France}
\affil{$^{26}$CNRS, IRAP, 14 avenue Edouard Belin, 31400 Toulouse, France}
\affil{$^{27}$European Southern Observatory (ESO), Alonso de Cordova 3107, Casilla 19001, Santiago 19, Chile}
\affil{$^{28}$McDonald Observatory, 16120 St Hwy Spur 78 \#2, Fort Davis, TX 79734, USA}
\affil{$^{29}$School of Math and Physics, University of Tasmania, Private Bag 37, GPO Hobart, 7001 Tasmania, Australia}
\affil{$^{30}$Physics Department, Faculty of Arts and Sciences, University of Rijeka, Omladinska 14, 51000 Rijeka, Croatia}
\affil{$^{31}$Technical University of Vienna, Department of Computing, Wiedner Hauptstrasse 10, Vienna, Austria}
\affil{$^{32}$Niels Bohr Institute, Astronomical Observatory, Juliane Maries vej 30, 2100 Copenhagen, Denmark}
\affil{$^{33}$Department of Physics \& Astronomy, San Francisco State University, 1600 Holloway Avenue, San Francisco, CA 94132, USA}
\affil{$^{34}$Perth Observatory, Walnut Road, Bickley, 6076 Perth, Australia}
\affil{$^{35}$University of the Free State, Faculty of Natural and Agricultural Sciences, Dept. of Physics, PO Box 339, 9300 Bloemfontein, South Africa}
\affil{$^{36}$South African Astronomical Observatory, PO Box 9, Observatory 7935, South Africa}
\affil{$^{37}$Astronomisches Rechen-Institut, Zentrum f\"ur Astronomie der Universit\"at Heidelberg (ZAH), M\"onchhofstr. 12-14, 69120 Heidelberg, Germany}
\affil{$^{38}$European Southern Observatory, Karl-Schwarzschild-Str. 2, 85748 Garching bei M\"unchen, Germany}
\affil{$^{39}$SUPA, School of Physics \& Astronomy, University of St Andrews, North Haugh, St Andrews KY16 9SS, UK}
\affil{$^{40}$Alsubai Establishment for Scientific Studies, Doha, Qatar}
\affil{$^{41}$Max Planck Institute for Solar System Research, Max-Planck-Str. 2, 37191 Katlenburg-Lindau, Germany}
\affil{$^{42}$Astrophysics Research Institute, Liverpool John Moores University, Liverpool CH41 1LD, UK}
\affil{$^{43}$The MOA Collaboration}
\affil{$^{44}$The $\mu$FUN Collaboration}
\affil{$^{45}$The PLANET Collaboration}
\affil{$^{46}$The RoboNet Collaboration}
\affil{$^{47}$Corresponding author}

\begin{abstract}
A planetary microlensing signal is generally characterized by a
short-term perturbation to the standard single lensing light curve. 
A subset of binary-source events can produce perturbations that
mimic planetary signals, thereby introducing an ambiguity between
the planetary and binary-source interpretations.
In this paper, we present
analysis of the microlensing event MOA-2012-BLG-486, for which the light
curve exhibits a short-lived perturbation. Routine modeling not
considering data taken in different passbands yields a best-fit
planetary model that is slightly preferred over the best-fit
binary-source model. However, when allowed for a change in the color
during the perturbation, we find that the binary-source model yields a
significantly better fit and thus the degeneracy is clearly resolved.
This event not only signifies the importance of considering various
interpretations of short-term anomalies, but also demonstrates the
importance of multi-band data for checking the possibility of
false-positive planetary signals.
\end{abstract}

\keywords{gravitational lensing -- binaries: general -- planetary systems}

\section{INTRODUCTION}
Microlensing is one of the major methods of detecting and characterizing
extrasolar planets. The method is important for the comprehensive
understanding of planet
formation because it is sensitive to planets that are
difficult for other methods to detect, especially planets near and
beyond the snow line 
\citep{2011Natur.473..349S, 2012Natur.481..167C, 2012ARA&A..50..411G}. 

It is known that such short-term perturbations in microlensing light
curves can be caused by scenarios other than planetary companions. In
particular, some subset of nearly equal-mass binary lenses can exhibit
perturbations with similar magnitudes and durations as those caused by
very low mass ratio companions 
\citep{2002ApJ...572.1031A, 2008ApJ...689...53H}. 
Another
example is when the lens is an isolated star, but the source is itself
a binary with a large flux ratio. In this case, if the lens passes
close to both members of the binary source, the light curve
can appear as a normal single lens curve, superposed with a short
duration perturbation
that results when the lens passes close to the fainter source 
\citep{1998ApJ...506..533G, 2004ApJ...611..528G}. 
\cite{1998ApJ...506..533G} 
pointed out that the frequency
of planetary-like perturbations produced by binary-source events can
be as high as planet-induced perturbations. Therefore, the analysis of
a short-term signal in a lensing light curve requires careful
examination considering all possible interpretations.

In this paper, we present our analysis of MOA-2012-BLG-486. 
The light curve of the event exhibits a short-term perturbation.
In \S\ 2, we describe the observation of the event. 
In \S\ 3, we describe the analyses conducted to examine 
the nature of the perturbation and present results. 
We summarize and conclude in \S\ 4.


\begin{deluxetable*}{lll}
\tablecaption{Telescopes\label{table:one}}
\tablewidth{0pt}
\tablehead{
\multicolumn{1}{c}{group}     &
\multicolumn{1}{c}{telescope} & 
\multicolumn{1}{c}{passband}  
}
\startdata
MOA       &  1.8m Mt. John Observatory, New Zealand                              &  MOA-Red        \\
$\mu$FUN  &  1.3m SMARTS, Cerro Tololo Inter-American Observatory (CTIO), Chile  &  $V$, $I$, $H$  \\
$\mu$FUN  &  0.6m Observatorio do Pico dos Dias (OPD), Brazil                    &  $I$            \\
PLANET    &  1.0m South African Astronomical Observatory (SAAO), South Africa    &  $I$            \\
RoboNet   &  2.0m Faulkes North Telescope (FTN), Hawaii, USA                     &  $I$            \\
RoboNet   &  2.0m Faulkes South Telescope (FTS), Australia                       &  $I$            
\enddata  
\tablecomments{ 
MOA-Red band is a custom wide band where the band width roughly 
corresponds to the sum of $R$ and $I$ bands.
}
\end{deluxetable*}

\section{OBSERVATIONS AND DATA}
The microlensing event MOA-2012-BLG-486 occurred on a star 
in the Galactic bulge field at the equatorial and Galactic coordinates 
(RA, DEC)$_{\rm J2000.0}$ = (18$^{\rm h}$01$^{\rm m}$08$^{\rm s}\hskip-2pt .$82, 
           $-$33$\arcdeg$13$\arcmin$06$\arcsec\hskip-2pt .$1) 
and ($l$, $b$)$_{\rm J2000.0}$ = (357.99$\arcdeg$, $-$5.03$\arcdeg$), respectively. 
It was discovered by
the Microlensing Observations in Astrophysics 
(MOA: \citealp{2001MNRAS.327..868B}; \citealp{2003ApJ...591..204S}) 
survey during the 2012 observing season. From the preliminary 
analysis of the event based on
the rising part of the light curve, it was found that the event would 
reach a high magnification.
Since such a high-magnification event is 
very sensitive to a planet \citep{1998ApJ...500...37G}, 
an alert was issued for intensive follow-up 
observations. Based on the 
alert, the event was additionally observed by 
other groups including Microlensing Follow-Up Network
($\mu$FUN: \citealp{2006ApJ...644L..37G}),  
Probing Lensing Anomalies NETwork 
(PLANET: \citealp{2006Natur.439..437B}), 
and RoboNet \citep{2009AN....330....4T}. 
In Table~\ref{table:one}, we list the survey and follow-up groups 
along with the telescopes used for the observations.
We note that although most data were taken in $I$ band,
some data sets were obtained in other bands.
Real-time modeling played and important role in the acquisition of
the color data that ultimately allowed us to distinguish between
binary-lens and binary-source models. The CTIO $VIH$ points over
the first peak were taken as part of normal $\mu$FUN observing strategy
(possible high-magnification event). But the later observations,
which measured the colors of the second peak were taken in direct
response to circulation of two models (binary-lens and
binary-source), in a specific effort to distinguish them.
The event did not return to its baseline brightness 
until the end of the 2012 bulge season.
For secure measurement of the baseline,
the event was additionally observed in 2013 season.

\begin{figure}[th]
\epsscale{1.15}
\plotone{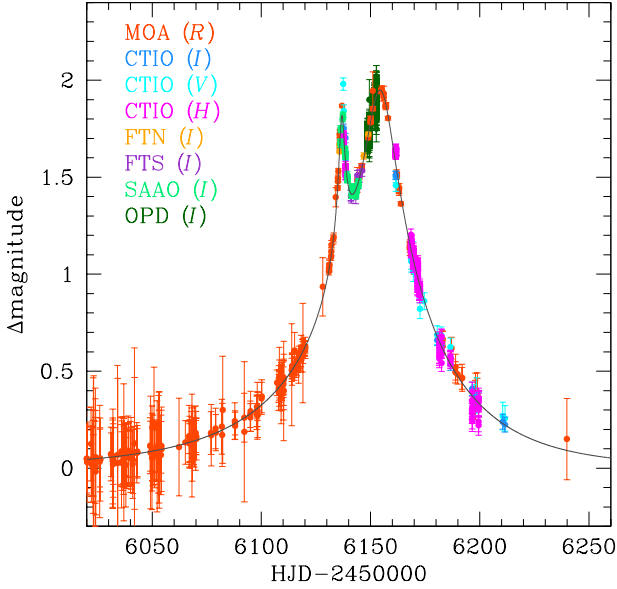}
\caption{\label{fig:one}
Light curve of MOA-2012-BLG-486. The presented model curve is based on the
single-band binary-source model. The notation in the
parenthesis after the legend of each observatory 
denotes the passband of observation. 
}
\end{figure}

Figure \ref{fig:one} shows the light curve of MOA-2012-BLG-486. 
It is characterized by a
short-term anomaly at HJD$'$ = HJD $-$ 2450000 $\sim$ 6137. 
Apart from the anomaly, 
the overall light curve is well described by the standard form of 
a single-lens event. Since the characteristics of the light curve strongly
indicate the possible existence of a planetary companion, we
conduct a detailed analysis of the event.

For the data sets used in the analysis, initial reductions were conducted
using photometry codes developed by the individual groups. For some data sets,
we conducted additional reduction to improve photometry using a pipeline
based on difference image analysis 
\citep{2001MNRAS.327..868B, 2009MNRAS.397.2099A}. 
Nevertheless, the photometric accuracy is limited because there exists a 
bright star close to the source.  
When this occurs, the photometric measurements are
often correlated with the seeing.
To minimize the seeing effect, 
we use data taken with seeing less than 3 arc-seconds.
We normalize
the photometry uncertainties of different data sets by first adding a
quadratic term so that the cumulative distribution of $\chi^2$ ordered by
magnification is approximately linear and subsequently rescaling the
errors so that $\chi^2$ per degree of freedom ($\chi^2/{\rm dof}$)
for each data set becomes unity
\citep{2012A&A...547A..55B, 2012ApJ...752...82M}.

\section{MODELING}
Knowing that a short-term perturbation in a lensing light curve can be
produced either by a binary (including planetary) 
companion to the lens or a binary companion to
the source, we conduct both binary-lens and binary-source modeling of the light
curve. Due to their physical natures, the two types of modeling require
widely different parameterization.

Binary lens modeling requires 7 basic parameters.
Among them, three parameters are needed to describe the lens-source approach, 
including the time of the closest lens-source approach, $t_0$, 
the lens-source separation at that moment, $u_0$ (impact parameter), 
and the time scale required for the source to cross the Einstein radius 
of the lens, $t_{\rm E}$ (Einstein time scale).
The Einstein ring represents the image of the source for the case of
exact lens-source alignment and its radius $\theta_{\rm E}$ (Einstein radius)
is commonly used as a length scale in lensing phenomena. 
We note that the lens-source impact parameter $u_0$
is normalized by $\theta_{\rm E}$.
Three additional parameters are needed to characterize the star-planet system,
including the projected separation, $s$ (normalized by $\theta_{\rm E}$),
the mass ratio, $q$, and the source trajectory angle with respect to the  
binary axis, $\alpha$. A planetary perturbation usually 
occurs when the source encounters a caustic,
and so finite-source effects become important. 
To account for these effects, one needs
the additional parameter $\rho_{\ast}=\theta_{\ast}/\theta_{\rm E}$ 
(normalized source radius),
where $\theta_{\ast}$ represents the angular radius of the source star.

\begin{figure}[th]
\epsscale{1.18}
\plotone{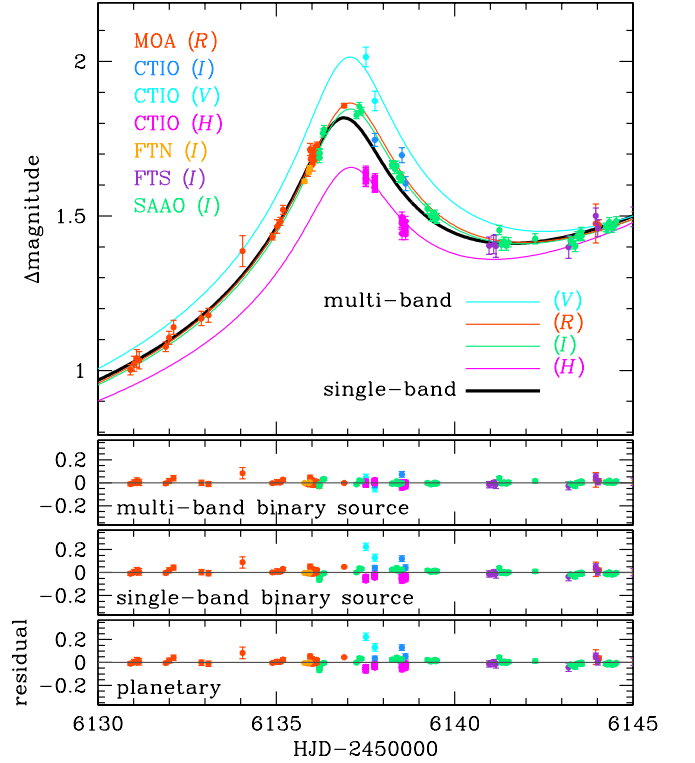}
\caption{\label{fig:two}
Zoom of the perturbation region of the light curve of MOA-2012-BLG-486. The
thick solid black curve is the best-fit model obtained from single-band
binary-source modeling, while thin curves with various colors are model curves
considering color effects. The 3 bottom panels show the residuals from the 
three sets of modeling.
}
\end{figure}

The lensing magnification of a binary-source event corresponds to the
flux-weighted mean of the single-lens magnifications associated with the
individual source stars, i.e. 
$A=(A_1F_{{\rm S},1}+A_2F_{{\rm S},2})/(F_{{\rm S},1}+F_{{\rm S},2})$
\citep{1992ApJ...397..362G}. 
Here $F_{{\rm S},i}$ and $A_i$ ($i=1,2$)
represent the flux and magnification for each source star. To describe the
lens approach to the individual source stars, one needs pairs of the impact
parameters ($u_{0,1}$ and $u_{0,2}$) and times of closest approach 
($t_{0,1}$ and $t_{0,2}$).
Furthermore, an additional parameter is needed to specify the flux ratio
between the source stars, $q_{\rm F}=F_{{\rm S},2}/F_{{\rm S},1}$.  
In our modeling, we set $F_{{\rm S},1}>F_{{\rm S},2}$ and thus
$q_{\rm F}<1.0$.

Besides the basic parameters, it is often needed to consider higher-order
effects to precisely describe lensing light curves. The event MOA-2012-BLG-486
lasted for two observing seasons. For such a long time-scale event, the
positional change of an observer caused by the Earth's orbital motion around
the Sun might cause a deviation in the lensing light curve due to
the deviation of the apparent lens-source motion from 
a rectilinear trajectory \citep{1992ApJ...392..442G}. 
Considering this parallax effect requires two additional lensing
parameters $\pi_{{\rm E},N}$ and $\pi_{{\rm E},E}$, 
which are the components of the lens parallax vector $\pivec_{\rm E}$
projected on the sky along the north and east equatorial axes,
respectively. For the binary-lens case, the positional change of the lens
caused by the orbital motion can also induce long-term deviations 
in lensing light curves. To first order approximation, 
the lens orbital effect is described
by the two parameters $ds/dt$ and $d\alpha/dt$ 
that represent the change rates of the normalized
binary separation and source trajectory angle, respectively 
\citep{2000ApJ...534..894A, 2002ApJ...572..521A}. 

\begin{deluxetable}{lrrr}
\tablecaption{Lensing Parameters\label{table:two}}
\tablewidth{0pt}
\tablehead{
\multicolumn{1}{c}{parameters} &
\multicolumn{1}{c}{planetary}  &
\multicolumn{2}{c}{binary source}  \\
\multicolumn{1}{c}{} &
\multicolumn{1}{c}{} &
\multicolumn{1}{c}{single band} &
\multicolumn{1}{c}{multi bands} 
}
\startdata
$\chi^2/{\rm dof}$         &  2828.8/2184          &  2855.7/2187          &  2165.3/2184         \\
$t_{0,1}$ (HJD$'$)         &  6149.33 $\pm$ 0.24   &  6154.43 $\pm$ 0.02   &  6154.24 $\pm$ 0.03  \\
$t_{0,2}$ (HJD$'$)         &           --          &  6137.19 $\pm$ 0.05   &  6137.20 $\pm$ 0.05  \\
$u_{0,1}$                  &    0.059 $\pm$ 0.002  &    0.099 $\pm$ 0.004  &    0.077 $\pm$ 0.003 \\
$u_{0,2}$                  &           --          &    0.024 $\pm$ 0.003  &    0.027 $\pm$ 0.002 \\
$t_{\rm E}$ (days)         &     92.8 $\pm$ 2.2    &     66.4 $\pm$ 1.9    &     78.1 $\pm$ 2.8   \\
$s$                        &     1.65 $\pm$ 0.01   &           --          &           --         \\
$q$                        &    0.029 $\pm$ 0.001  &           --          &           --         \\
$\alpha$                   &    2.789 $\pm$ 0.006  &           --          &           --         \\
$\rho_{\ast}$ ($10^{-3}$)  &      0.5 $\pm$ 0.3    &           --          &           --         \\
$\pi_{{\rm E},N}$          &    -0.51 $\pm$ 0.04   &    -0.18 $\pm$ 0.11   &    -0.37 $\pm$ 0.09  \\
$\pi_{{\rm E},E}$          &     0.11 $\pm$ 0.03   &     0.17 $\pm$ 0.03   &     0.08 $\pm$ 0.02  \\
$ds/dt$                    &    -1.62 $\pm$ 0.28   &           --          &           --         \\
$d\alpha/dt$               &    -0.46 $\pm$ 0.13   &           --          &           --         \\
$q_{F}$                    &           --          &    0.097 $\pm$ 0.002  &           --         \\
$q_{{\rm F},V}$            &           --          &           --          &    0.163 $\pm$ 0.007 \\
$q_{{\rm F},R}$            &           --          &           --          &    0.125 $\pm$ 0.003 \\
$q_{{\rm F},I}$            &           --          &           --          &    0.118 $\pm$ 0.003 \\
$q_{{\rm F},H}$            &           --          &           --          &    0.081 $\pm$ 0.002  
\enddata                             
\tablecomments{ 
HJD$'$=HJD-2450000.
}
\end{deluxetable}

For both binary-lens and binary-source models, we look for best-fit sets of
lensing parameters by running a Markov Chain Monte Carlo search of
parameter space. In the initial binary-source modeling, we model the light
curve with a single flux ratio (single-band model). In Table~\ref{table:two}, 
we list the best-fit parameters for the individual models. 
We find that the overall
shape and the short-term feature of the light curve can be described by both
planetary and binary-source models with similar values of $\chi^2$:
$\chi^2=2828.8$ for the planetary model and
$\chi^2=2855.7$ for the binary-source model (single band). 
However, there exist some residuals from the models, 
especially in the region of the short-term anomaly as shown 
in the lower panels of Figure~\ref{fig:two}. 
Although consideration of parallax and orbital (for the planetary model) effects
somewhat improves the fits of both models, they contribute
to the fits of the overall shape of the light curve, not to the anomaly
feature. This indicates the need to consider another higher-order effect.

\begin{figure}[th]
\epsscale{1.15}
\plotone{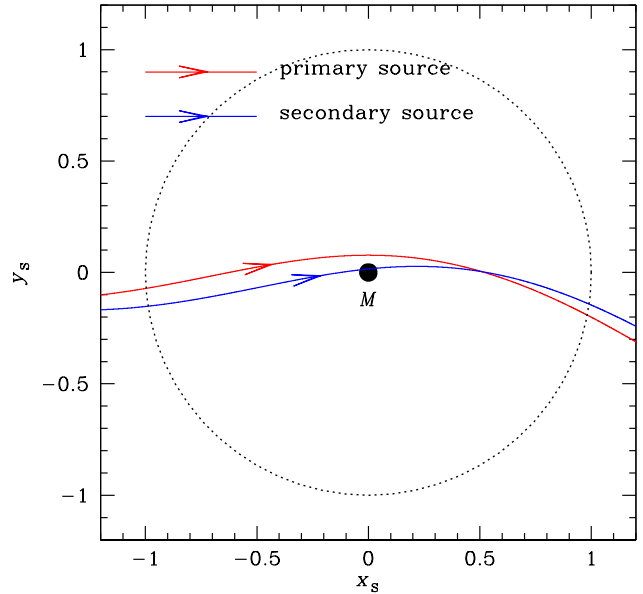}
\caption{\label{fig:three}
Geometry of the lens system for the binary-source model. The lens is
positioned at the origin (marked by $M$) and the dotted circle is
the Einstein ring. The 2 curves with arrows
represent the trajectories of the two source stars.
Lengths are scaled by the Einstein radius.
}
\end{figure}

\begin{figure}[th]
\epsscale{1.15}
\plotone{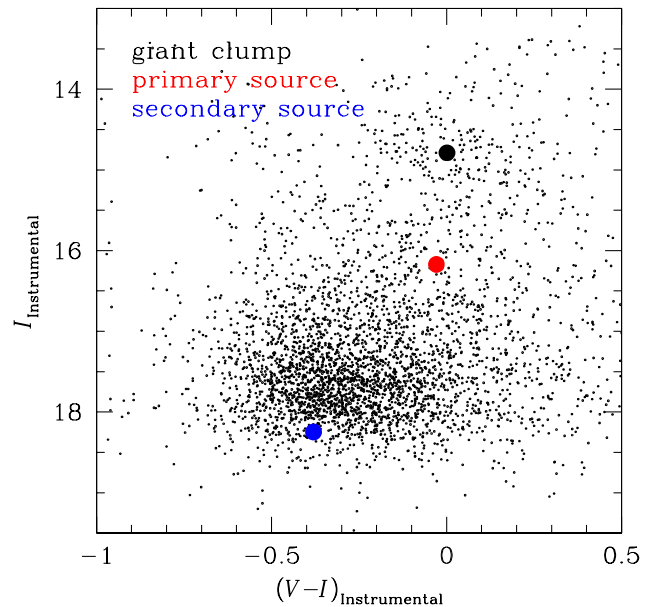}
\caption{\label{fig:four}
Positions of the components stars of the binary source 
in the color-magnitude diagram.
}
\end{figure}

Color change can occur for a binary-source event. This color
change occurs because the stars comprising a binary source have, in general,
different colors and thus the colors measured at the moments of the lens' 
approaches close to the individual source stars can be different. From the
initial single-band binary-source modeling, the measured flux ratio between
the two source stars, $q_{\rm F}\sim0.1$, is small.
This implies that the types of the
source stars are likely to be very different and thus color effects would be
pronounced during the perturbation. 
To check whether the color effect can explain the residual in the anomaly, 
we conduct additional binary-source modeling.
For this, we specify flux ratios for the
individual passbands of the data covering the anomaly, including
$V$ (CTIO), $R$ (MOA), $I$ (CTIO, FTN, FTS, SAAO), and $H$ (CTIO) band.

In Figure~\ref{fig:two}, we present the light curves in the region of the anomaly
resulting from modeling considering the color effect 
(multi-band binary-source model). 
We note that since
lensing magnifications depend on the observed passband, 4 model light
curves corresponding to the individual passbands are presented.
The measured flux ratio of each passband between the binary source stars
is listed in Table~\ref{table:two}. We find that the model considering the color
effect significantly improves the fit with $\Delta\chi^2=690.4$ compared
to the single-band binary-source model. Furthermore, the multi-band
binary-source model is better than the planetary model
by $\Delta\chi^2=663.5$. The improvement of the fit can be
visually seen from the residuals presented in Figure~\ref{fig:two}.
The superiority of the fit compared to the planetary model
combined with the obvious color effect strongly supports the
binary-source interpretation of the anomaly.

In Figure~\ref{fig:three}, we present the trajectories of the individual source stars
with respect to the lens for the best-fit binary-source model. 
The fact that the event is brighter in shorter wavelength passbands 
during the anomaly indicates that the faint
source component (secondary) is bluer than the bright component (primary). 
We choose the color
of each source trajectory in order to reflect the color of each star.
According to the model, the faint blue star approached very close to the
lens, producing a short-term anomaly, and then the bright red star followed.
The flux ratio between source stars is small, 
ranging from 0.08 ($H$ band) to 0.16 ($V$ band). 
As a result, the light curve is dominated by the light from the
bright source star except for the short term of the lens' approach 
to the faint source star.
We note that the trajectories are curved due to parallax effects.

In Figure~\ref{fig:four}, we present the locations of the binary source
components in the color-magnitude diagram. Based on the source
flux measured from modeling combined with the reference
position of the centroid of clump giants with the known
de-reddened magnitude at the Galactic distance $I_0=14.54$
\citep{2013ApJ...769...88N} 
and color $(V-I)_0=1.06$
\citep{2011A&A...533A.134B}, 
we determine the de-reddened magnitude and color of the individual
source stars as $(I, V-I)_0=(15.93, 1.03)$ for the primary and
$(18.25, 0.68)$ for the secondary.
These correspond to a K-type giant and a turn-off star
for the primary and secondary, respectively.

We check the possibility of the limb-darkening surface brightness
variation of the source star as a cause of the color variation.
From additional modeling, we find that $\chi^2$ improvement by the
limb-darkening effect is negligible and thus exclude the possibility.

\section{CONCLUSION}
We analyzed the microlensing event MOA-2012-BLG-486 where the light curve
exhibited a short-term perturbation indicating the 
possible existence
of a planetary companion to the lens. By conducting
detailed modeling of the light curve considering both planetary and
binary-source interpretations, we found that the perturbation was better
explained by the binary-source model. The degeneracy
between the planetary and binary-source interpretations was clearly
resolved via the color effect that occurred during the anomaly
thanks to the multi-band data obtained during the anomaly. 
The event not only signifies the importance of considering
various interpretations of short-term anomalies but also
demonstrates the importance of multi-band data for checking
the possibility of false-positive planetary signals.

Our ability to distinguish between the planetary and 
binary-source solutions rested critically on $VIH$ photometry.
However, while such dense multi-band photometry is routine
for follow-up observations from $\mu$FUN SMARTS CTIO,
microlens survey observations are typically taken overwhelmingly
in one band. With the advent of second generation surveys, a 
large fraction of microlensing planets are being detected from
survey-only data, and this will be increasingly true in the future.
Hence, the protocols for these surveys should be carefully
evaluated to make sure that color coverage is adequate to
distinguish binary-source events from planetary events.\\

\acknowledgments
Work by CH was supported by Creative Research Initiative 
Program (2009-0081561) of National Research Foundation of Korea.
The MOA experiment was supported by grants JSPS22403003 and JSPS23340064.
TS acknowledges the support JSPS 24253004. 
TS is supported by the grant JSPS23340044. 
YM acknowledges support from JSPS grants JSPS23540339 and JSPS19340058.
AG and BSG acknowledge support from NSF AST-1103471.
BSG, AG, and RWP acknowledge support from NASA grant NNX12AB99G.
SD was supported through a Ralph E. and Doris M. Hansmann
Membership at the IAS and NSF grant AST-0807444.
DMB, MD, KH, MH, SI, CS, RAS and YT are supported by 
NPRP grant NPRP-09-476-1-78 from the Qatar National Research Fund (a
member of Qatar Foundation). 
KH is supported by a Royal Society Leverhulme Senior Research Fellowship.
KH is supported by a Royal Society Leverhulme Trust Research Fellowship.
CS received funding from the European Union Seventh Framework Programme
(FP7/2007-2013) under grant agreement no. 268421.
The research leading to these results has received funding from the
European Community's Seventh Framework Programme
(/FP7/2007-2013/) under grant agreement No 229517.


\begin{thebibliography}{99}

\bibitem[Albrow et al.(2002)]{2002ApJ...572.1031A}
Albrow, M.~D., An, J., Beaulieu, J.-P., et al.\ 2002, \apj, 572, 1031 

\bibitem[Albrow et al.(2000)]{2000ApJ...534..894A} 
Albrow, M.~D., Beaulieu, J.-P., Caldwell, J.~A.~R., et al.\ 2000, \apj, 534, 894 

\bibitem[Albrow et al.(2009)]{2009MNRAS.397.2099A} 
Albrow, M.~D., Horne, K., Bramich, D.~M., et al.\ 2009, \mnras, 397, 2099 

\bibitem[An et al.(2002)]{2002ApJ...572..521A} 
An, J.~H., Albrow, M.~D., Beaulieu, J.-P., et al.\ 2002, \apj, 572, 521 

\bibitem[Bachelet et al.(2012)]{2012A&A...547A..55B} 
Bachelet, E., Fouqu{\'e}, P., Han, C., et al.\ 2012, \aap, 547, A55

\bibitem[Beaulieu et al.(2006)]{2006Natur.439..437B} 
Beaulieu, J.-P., Bennett, D.~P., Fouqu{\'e}, P., et al.\ 2006, \nat, 439, 437 

\bibitem[Bensby et al.(2011)]{2011A&A...533A.134B}
Bensby, T., Ad{\'e}n, D., Mel{\'e}ndez, J., et al.\ 2011, \aap, 533, A134 

\bibitem[Bond et al.(2001)]{2001MNRAS.327..868B} 
Bond, I.~A., Abe, F., Dodd, R.~J., et al.\ 2001, \mnras, 327, 868 

\bibitem[Cassan et al.(2012)]{2012Natur.481..167C}
Cassan, A., Kubas, D., Beaulieu, J.-P., et al.\ 2012, \nat, 481, 167 

\bibitem[Gaudi(1998)]{1998ApJ...506..533G} 
Gaudi, B.~S.\ 1998, \apj, 506, 533 

\bibitem[Gaudi(2012)]{2012ARA&A..50..411G} 
Gaudi, B.~S.\ 2012, \araa, 50, 411 

\bibitem[Gaudi \& Han(2004)]{2004ApJ...611..528G} 
Gaudi, B.~S., \& Han, C.\ 2004, \apj, 611, 528 

\bibitem[Gould(1992)]{1992ApJ...392..442G} 
Gould, A.\ 1992, \apj, 392, 442 

\bibitem[Gould et al.(2006)]{2006ApJ...644L..37G} 
Gould, A., Udalski, A., An, D., et al.\ 2006, \apjl, 644, L37 

\bibitem[Griest \& Hu(1992)]{1992ApJ...397..362G} 
Griest, K., \& Hu, W.\ 1992, \apj, 397, 362 

\bibitem[Griest \& Safizadeh(1998)]{1998ApJ...500...37G} 
Griest, K., \& Safizadeh, N.\ 1998, \apj, 500, 37 

\bibitem[Han \& Gaudi(2008)]{2008ApJ...689...53H} 
Han, C., \& Gaudi, B.~S.\ 2008, \apj, 689, 53 

\bibitem[Miyake et al.(2012)]{2012ApJ...752...82M} 
Miyake, N., Udalski, A., Sumi, T., et al.\ 2012, \apj, 752, 82 

\bibitem[Nataf et al.(2013)]{2013ApJ...769...88N} 
Nataf, D.~M., Gould, A., Fouqu{\'e}, P., et al.\ 2013, \apj, 769, 88 

\bibitem[Sumi et al.(2003)]{2003ApJ...591..204S} 
Sumi, T., Abe, F., Bond, I.~A., et al.\ 2003, \apj, 591, 204 

\bibitem[Sumi et al.(2011)]{2011Natur.473..349S} 
Sumi, T., Kamiya, K., Bennett, D.~P., et al.\ 2011, \nat, 473, 349 

\bibitem[Tsapras et al.(2009)]{2009AN....330....4T} 
Tsapras, Y., Street, R., Horne, K., et al.\ 2009, Astronomische Nachrichten, 330, 4 

\end{thebibliography}
\end{document}